\documentclass{amsart}

\usepackage{enumerate}
\usepackage{amssymb}
\usepackage{epsfig}
\usepackage{subfigure}

\newcommand{\dfn}{\textbf} 
\newcommand{\mdfn}[1]{\dfn{\mathversion{bold}#1}}

\newtheorem{theorem}{Theorem}[section]
\newtheorem{thm}[theorem]{Theorem}

\newtheorem{prop}[theorem]{Proposition}
\newtheorem{conjecture}[theorem]{Conjecture}

\theoremstyle{definition}
\newtheorem{defn}[theorem]{Definition}
\newtheorem{example}[theorem]{Example}

\theoremstyle{remark}
\newtheorem{remark}[theorem]{Remark}

\numberwithin{equation}{section}

\newcommand{\sse}{\subseteq}

\newcommand{\spse}{\supseteq}
\newcommand{\sps}{\supset}
\newcommand{\power}{\mathcal{P}}
\newcommand{\finite}{\mathcal{P}'}
\newcommand{\st}{\mid}
\newcommand{\N}{\mathbb{N}}
\newcommand{\ra}{\rightarrow}
\newcommand{\Path}{\mathrm{Path}}
\newcommand{\Fut}{\mathrm{Fut}}
\newcommand{\Past}{\mathrm{Past}}
\newcommand{\future}[1]{\left.#1\right\uparrow}
\newcommand{\subfuture}[1]{\left.#1\right\Uparrow}
\newcommand{\past}[1]{\left.#1\right\downarrow}
\newcommand{\subpast}[1]{\left.#1\right\Downarrow}
\newcommand{\Inc}{\mathrm{Inc}}

\newcounter{en}
\newenvironment{enum}{\begin{list}{\arabic{en}.}
{\usecounter{en} \setlength{\leftmargin}{20pt} \setlength{\itemsep}{8pt}
\setlength{\topsep}{5pt}}}
{\end{list}}

\newcounter{ena}
\newenvironment{enuma}{\begin{list}{\alph{ena})}
{\usecounter{ena} \setlength{\leftmargin}{20pt} \setlength{\itemsep}{2pt}
\setlength{\topsep}{5pt}}}
{\end{list}}

\begin{document}

\title{Causal Sites as Quantum Geometry}

\author{J. Daniel Christensen}
\address{Department of Mathematics \\
University of Western Ontario \\
London, Ontario, Canada }
\email{jdc@uwo.ca}

\author{Louis Crane}
\address{Mathematics Department\\
Kansas State University\\
Manhattan, KS 66506 USA}
\email{crane@math.ksu.edu}

\date{28 September 2005}

\begin{abstract}
We propose a structure called a causal site to use as a setting for
quantum geometry, replacing the underlying point set. The structure
has an interesting categorical form, and a natural ``tangent
2-bundle,'' analogous to the tangent bundle of a smooth
manifold. Examples with reasonable finiteness conditions have an
intrinsic geometry, which can approximate classical solutions to
general relativity. We propose an approach to quantization of causal
sites as well.
\end{abstract}

\maketitle

\section{Introduction and Physical Motivation}

This paper is part of a program to found quantum gravity in relational
topology; more precisely, to replace point set topology with a special
type of category as the underlying structure on which to put
geometrodynamics.

Physically, the idea is that what we actually observe are interactions
between bounded regions of spacetime. These could be either material
systems or regions of empty space whose causal effects can be directly
or indirectly distinguished by material systems. There should be a
direct mathematical description of the flow of information between the
regions, and points should appear only relative to an observer, as
minimal distinguished regions. We will propose a specific axiomatic
structure for a type of category which would contain the regions as
objects and the relationships between them as different types of
morphisms.
The hope is that this would lead to a description of quantum physics
free from ultraviolet divergences, by eliminating the underlying point
set continuum.

Since categories as generalizations of topological spaces
are well known in mathematics, where they are referred to as sites
\cite{Ar, MM}, we are calling our new structures \dfn{causal sites}.
Strictly speaking, a site is a category together with a structure
called a ``Grothen\-dieck topology,'' which is the analog of a
topology for an ordinary space. We will explain below that our axioms
also involve a structure very similar to a Grothendieck topology.

This paper will mainly discuss the ``topological'' side of the
problem, i.e.\ the structure of the causal sites themselves. We
discover that, unlike manifolds, causal sites with suitable finiteness
conditions have an intrinsic geometry. Thus the distinction between
topology and geometry is bridged over in our new picture. Examples are
described which reproduce the classical solutions to general
relativity above the Planck scale. We briefly consider the possibility
of directly quantizing causal sites, thus directly producing a quantum
theory of general relativity.

We will then briefly consider the appropriate classes of presheaves
over causal sites, namely unitary and bisimplicial prestacks.  We hope
this can serve as a bridge between the topological aspects of causal
site theory and the problem of constructing quantum physics over
them. We will make some remarks as to how to construct a model for
quantum gravity at the end.

A very important feature of the topology of causal sites is that they
have a tangent 2-bundle, which is analogous to the tangent bundle of a
manifold. We believe this will serve as a setting for applications to
quantum geometry and physics.

The structure of \dfn{causal sites} is a synthesis of two
constructions, one well known from homotopy theory and algebraic
geometry, the other familiar in the relativity literature. A
\dfn{site}~\cite{Ar} is a category thought of as a generalization of
the lattice of open sets of a topological space, with a distinguished
family of covers for objects. (We warn the reader that our construction
does not fully conform to the accepted definition of a site, since the 
covers in it do not satisfy the axioms of a Grothendieck topology, as
 explained below.)  The morphisms of the category represent
an abstract version of containment. Doing sheaf theory over such
generalized spaces is an important part of modern mathematics.

On the other hand, a \dfn{causal set}~\cite{So, HMS}, or partially
ordered set, is a discrete point set approximating the causal
structure of a spacetime manifold.

Now, since up to this point nobody has tried to use sites as a
foundation for relativity, the natural structure that occurs when we
combine the two ideas has never been considered. We will show that it
is surprisingly rich and elegant.

Since causal sets are well known as models for general relativity, let us 
mention some of their similarities and differences with causal sites.
Causal sites with a suitable finiteness condition have an intrinsic metric 
structure, as mentioned above. It is defined by counting the length of a 
maximal causal chain.  This is quite similar to the intrinsic metric of a 
discrete causal set~\cite{So}. We believe the greater flexibility of the 
families of 
regions in a causal site will mean they give much better approximations to 
the geometry of spacetimes than causal sets do. There can be infinitely 
many regions intermediate between two given ones in a causal site, and yet
causal paths can have a maximal finite length, as we show in examples below.
While random causal sets which approximate Minkowski space are believed to 
reproduce Lorentz invariance in the infinite volume limit, examples of 
causal sites which approximate Minkowski space have an invariance under
germs of the Lorentz group on bounded regions as well. As models for general 
relativity, causal sites can be studied in many of the same ways as causal 
sets. Mathematically, causal sites are a ``categorification'' of causal sets.

The fundamental mathematical observation is that both the containment
structure of a site and the causal structure of a causal set can be
described as partial orders. As we will explain in 
Section~\ref{se:simplicial}, these two partial orders give
rise to a natural bisimplicial set which we call the 
elementary classifying space of the causal site.  From 
this point of view, certain collections of regions can be 
viewed as products of simplices, and these products are glued together 
in a well-defined way.

We also construct another bisimplicial set, which we call the
physical classifying space.  This turns out to be a ``special''
bisimplicial set, a type of bisimplicial set which corresponds
to a weak 2-category~\cite{Ta}. 
This means that causal sites fall into a class of structure
which is already studied, and that the family of simplicial presheaves
over it is well understood.

We will argue at the end that this suggests an approach to doing
quantum general relativity over causal sites, by putting state sum
models on their tangent 2-bundles.

This paper is meant to open a number of lines of research. The
contents are as follows: in Section 2 we present the axioms for a
causal site and some simple consequences. In Section 3 we show how a
causal site with suitable finiteness conditions can have an intrinsic
geometry.
In Section 4, we briefly discuss quantization of sites and in Section 5 
we discuss the simplicial and bisimplicial structures inherent in a 
causal site. In Section 6,
we translate this into a bicategorical structure. Sections 7 and 8
discuss the structure of the tangent spaces and tangent bundle, while
Section 9 discusses the general analog of a bundle for a causal site, namely a
prestack. In Sections 10 and 11 we discuss how state sum models could emerge in
a causal site. Section 12 contains conclusions and outlook.

\subsection{Categorical background}

We freely make use of terminology and ideas from category theory.
In particular, we assume basic familiarity with weak 2-categories,
which are also called bicategories.  Good references for this
topic are~\cite[Ch.\ 1]{Le04}, \cite{Le98} and~\cite{Be}.  
We also give more specific references as needed.

\section{The Axiomatic Structure of Causal Sites}

We are now going to axiomatize the structure of regions in a causal
spacetime.

\begin{example}\label{ex:lorentz}
Let $M$ be a Lorentzian manifold with no closed timelike curves and a
global time orientation.  
For points $p$ and $r$ in $M$, write $p \ll r$ if there is a
future-directed timelike curve from $p$ to $r$,
and let $D(p,r)$ be the set of all points $q$ with $p \ll q \ll r$.
We call $D(p,r)$ a \dfn{diamond}, and we say that
a subset $A$ of $M$ is \dfn{bounded} if it is contained in a finite
union of diamonds.  For $A$ and $B$ bounded regions, write 
$A \sse B$ when $A$ is a subset of $B$, and write $A \prec B$ when
every point in region $A$ is in the causal past of every point in
region $B$.  That is, for each $a$ in $A$ and $b$ in $B$, $a \ll b$.

The motivation for the above definition of ``bounded'' is the following.
If a subset $A$ has compact closure, then it is bounded in the above sense.
And if the manifold $M$ is globally hyperbolic, then the converse holds.
So for globally hyperbolic manifolds, bounded is equivalent to
compact closure.  However, in general, our completion axiom below
is only satisfied if one allows regions without compact closure.
\end{example}

Below we list some properties that this set of regions has.  We then
want to consider more general systems satisfying the axioms. We
believe the interesting examples will actually have fewer regions than
the example above coming from a Lorentzian manifold.

\begin{defn}
A \dfn{causal site} is a set of ``regions'' with two binary relations
denoted $\sse$ and $\prec$ satisfying the axioms below.
If $A \sse B$ we say that $A$ is a \dfn{subset} of $B$ or that
$B$ \dfn{contains} $A$.  If $A \prec B$ we say that $A$ \dfn{precedes} $B$.

\begin{enum}
\item $\sse$ is a partial order on the set of regions.
This means that for all regions $A$, $B$ and $C$:
  \begin{enuma}
  \item $A \sse B$ and $B \sse C$ implies $A \sse C$;
  \item $A \sse A$;
  \item $A \sse B$ and $B \sse A$ implies $A = B$.
  \end{enuma}

\item The partial order $\sse$ has a minimum element $\phi$.
This means that $\phi$ is contained in every region $A$.  
This uniquely determines $\phi$, and $\phi$ is called the
\dfn{empty region}.

\item The partial order $\sse$ has \dfn{unions}.
This means that for all regions $A$ and $B$, there exists
a region $A \cup B$ such that:
  \begin{enuma}
  \item $A \sse A \cup B$ and $B \sse A \cup B$;
  \item if $A \sse C$ and $B \sse C$ then $A \cup B \sse C$.
  \end{enuma}
These requirements uniquely determine $A \cup B$, and the
binary operation $\cup$ is automatically associative and
commutative.
 
\item $\prec$ induces a strict partial order on the non-empty regions.
This means that for all non-empty regions $A$, $B$ and $C$:
  \begin{enuma}
  \item $A \prec B$ and $B \prec C$ implies $A \prec C$;
  \item $A$ does not precede $A$.
  \end{enuma}

\item For all regions $A$, $B$ and $C$,
$A \sse B$ and $B\prec C$ implies $A \prec C$.

\item For all regions $A$, $B$ and $C$,
$A \sse B$ and $C \prec B$  implies $C \prec A$.

\item For all regions $A$, $B$ and $C$,
$A \prec C$ and $B \prec C$ implies $A \cup B \prec C$.

\item \label{ax:cutting}
For all regions $A$ and $B$, there exists a region 
$B_A$ such that:
  \begin{enuma}
  \item $B_A \prec A$ and $B_A \sse B$;
  \item if $D \prec A$ and $D \sse B$ then $D \sse B_A$.
  \end{enuma}
These requirements uniquely determine $B_A$, and $B_A$ is
called the \dfn{cutting} of $A$ by $B$.  Note that $B_A$
can be empty.

\item \label{ax:completion}
If $A$ and $C$ are non-empty regions such that $A \prec C$
and there exists a $D$ with $A \prec D \prec C$, 
then there exists a $B$ complete with respect to $A \prec C$.
The definition of ``complete'' is below (Definition~\ref{de:completion}).
\end{enum}
\end{defn}

\begin{defn}
Regions $A$ and $B$ in a causal site are \dfn{disjoint} if the
only region which is contained in both $A$ and $B$ is the empty region.
More generally, a set of regions is \dfn{disjoint} if each pair
of regions it contains is disjoint.
\end{defn}

Note that if $A \prec B$ then $A$ and $B$ are disjoint.

\begin{defn}
Suppose $S$ and $T$ are sets of disjoint regions.
If every region in $S$ contains some region in $T$, we
say that $T$ is a \dfn{refinement} of $S$.
That is, $T$ is obtained from $S$ by shrinking
some regions and adding new regions.

A \dfn{causal path} $P$ is a sequence $A_1 \prec A_2 \prec \cdots \prec A_m$
of non-empty regions.
If $A \prec A_1$ and $A_m \prec B$, we say that $P$ is a causal path
\mdfn{from $A$ to $B$} and has \dfn{length} $m+1$.
There is exactly one causal path of length $1$.

A \dfn{refinement} of the causal path 
$A_1 \prec A_2 \prec \cdots \prec A_m$
is a causal path 
$C_1 \prec C_2 \prec \cdots \prec C_n$
such that $\{C_j\}$ is a refinement of $\{A_i\}$.
\end{defn}

If $P$ is a causal path from $A$ to $B$, and $A' \prec A$, then $P$ is
also a causal path from $A'$ to $B$.  Analogous statements can be
made when $A' \sse A$, $B \prec B'$ and $B' \sse B$.
See Section~\ref{se:2-bundle} for more details.


\begin{defn}\label{de:completion}
If $A \prec B \prec C$, we say that $B$ is \dfn{complete} for the 
causal pair $A \prec C$ if every causal path
from $A$ to $C$ can be refined to a causal path from $A$ to $C$
one of whose members is contained in $B$.
$B$ is called a \dfn{completion} of $A \prec C$.
\end{defn}

\begin{figure}
\centering
\mbox{\subfigure[Completion]{\epsfig{file=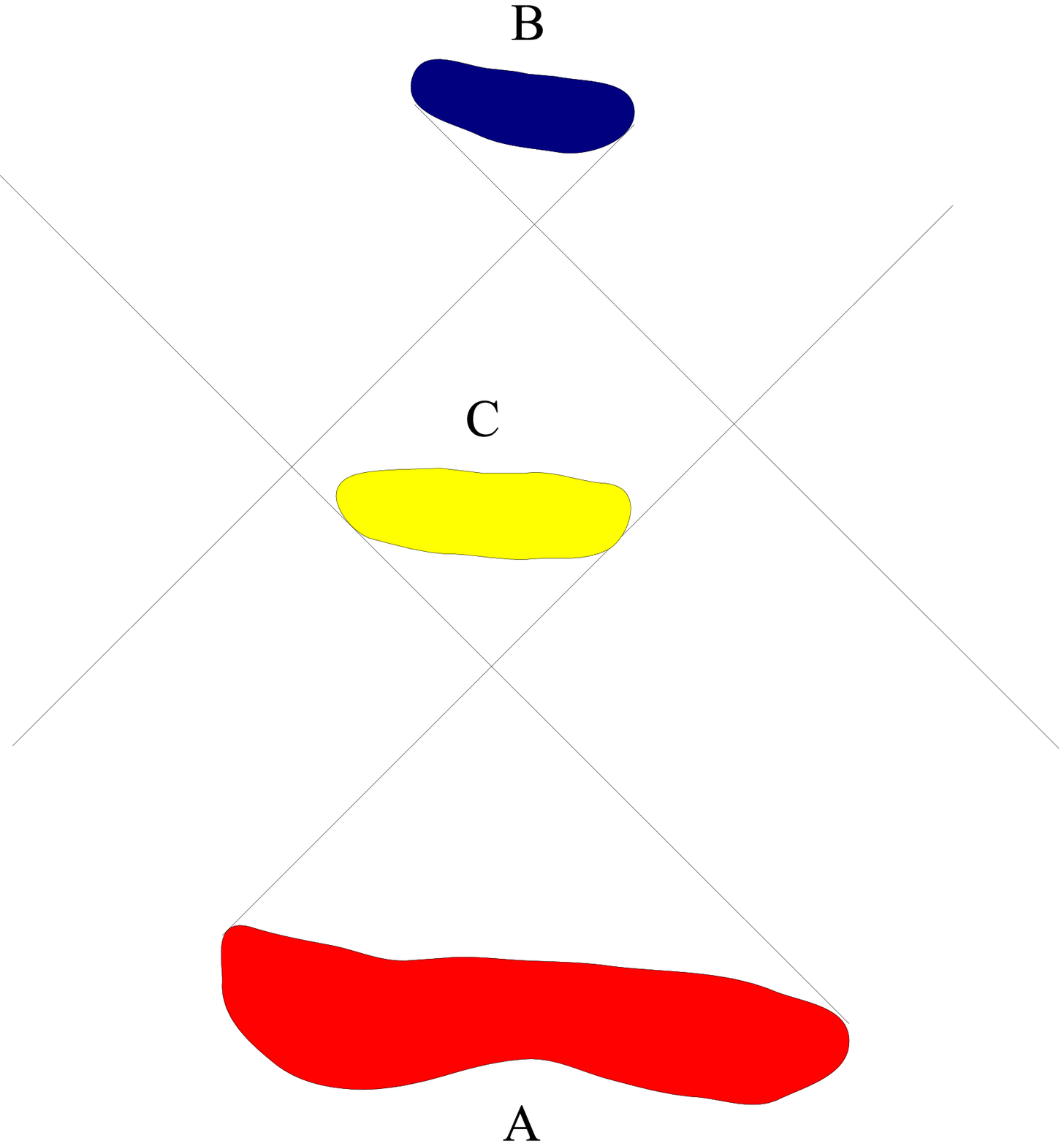,height=2in}}
\qquad\subfigure[Strong completion]{\epsfig{file=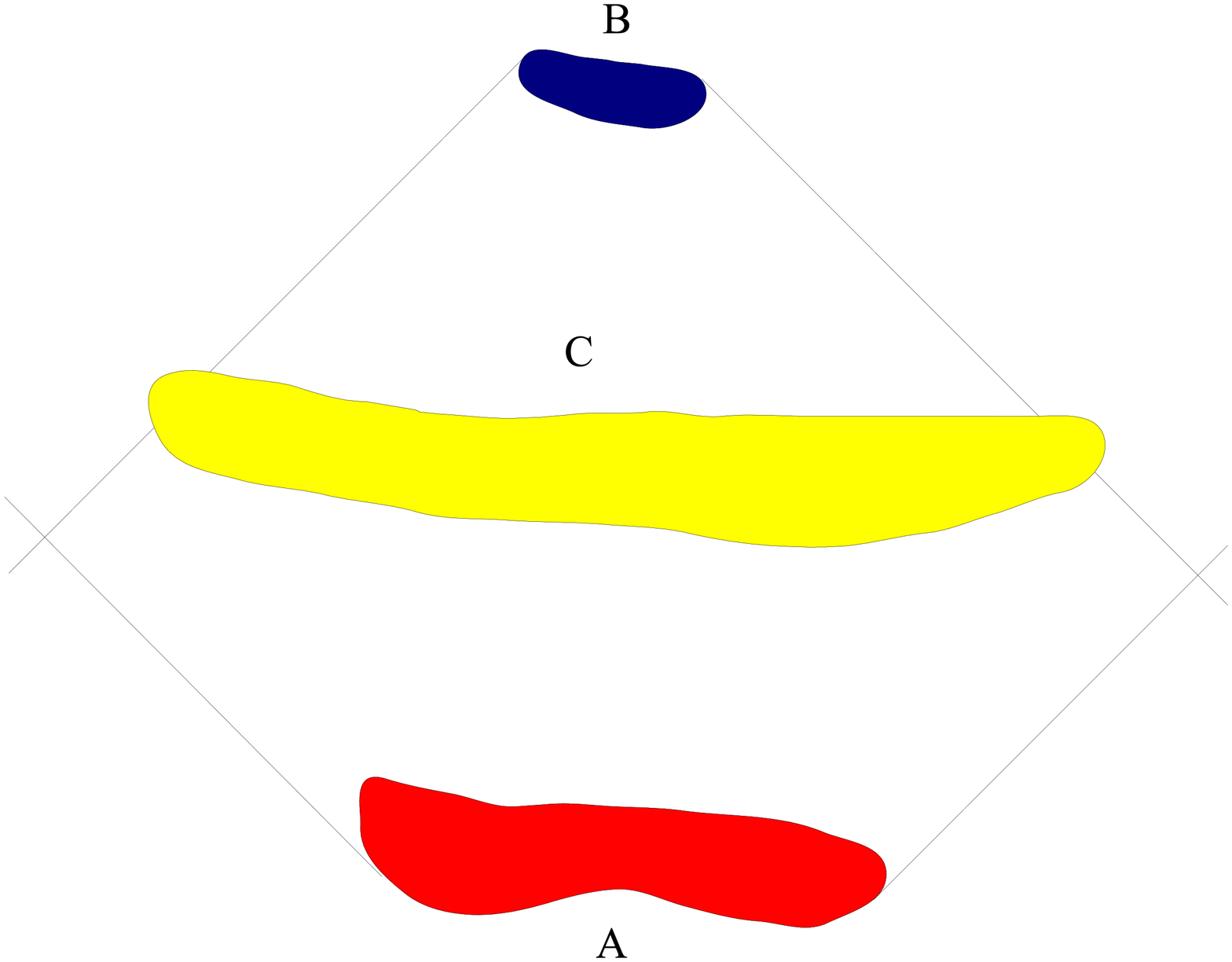,height=2in}}}
\caption{Comparing completions}
\label{fig:completions}
\end{figure}

See Figure~\ref{fig:completions} for an example of a completion.
The final axiom of a causal site requires that completions exist.
But note that they are rarely unique.

There are some elementary consequences whose proofs we leave to the
reader.

\begin{prop}
The following are true in any causal site.
\begin{enum}
\item For every region $A$, $\phi \prec A$ and $A \prec \phi$.

\item $B_A = B$ iff $B \prec A$.

\item $B \sse C$ implies $B_A \sse C_A$.

\item $B_A \cup C_A \sse (B \cup C)_A$.  (The reverse inclusion
fails for some of our examples.)

\item The collection of regions of a causal site which
precede a region $B$ forms another causal site. 
When $B$ is non-empty, 
this causal site is called the \dfn{local site} of $B$.
\end{enum}
\end{prop}

\begin{example}\label{ex:lorentz-II}
Let $M$ be as in Example~\ref{ex:lorentz}.  Then the bounded regions
in $M$, with the relations $\sse$ and $\prec$ defined earlier, satisfy
the axioms. 

We'll check the non-trivial axioms.
For the cutting axiom (Axiom~\ref{ax:cutting}),
let $B_A$ be the set of all points $p$ such that
$\{p\} \prec A$ and $p \in B$.
This is bounded (contained in a finite union of diamonds) since $B$ is.

For the completion axiom (Axiom~\ref{ax:completion} and
Definition~\ref{de:completion}),
a completion of $A \prec C$ can be taken to be the set of all
points $p$ such that $A \prec \{p\} \prec C$.
This is bounded because it is a subset of the diamond
$D(a,c)$ for any points $a \in A$ and $c \in C$.
(In general, this completion may not have compact closure.)

The remaining axioms are straightforward.
\end{example}

\begin{example} \label{ex:power}
Let $(M,\leq)$ be any poset (partially ordered set) and 
let $\power(M)$ be the set of
all subsets of $M$.  We define $\sse$ to be the usual subset
relation, and for $A$ and $B$ subsets of $M$, we say that
$A \prec B$ iff $\forall a \in A \,\, \forall b \in B \,\, a < b$.
(Note that we use the strict inequality $a < b$.)
Then $\power(M)$, $\sse$, and $\prec$ form a causal site.
A completion of $A \prec B$ can be taken to be
$\{m \in M \st \forall a \in A \,\, \forall b \in B \,\, a < m < b\}$.
\end{example}

\begin{example} \label{ex:finite}
Let $(M,\leq)$ be a poset such that $\{n \in M \st m \leq n \leq m'\}$
is finite for each $m$ and $m'$ in $M$.
(Such a poset is called \dfn{locally finite} or a \dfn{causal set}~\cite{So}.)
Let $\finite(M)$ be the set of finite subsets of $M$.
Then $\finite(M)$ is a causal site using relations defined as in the
previous example.  The hypothesis on $M$ ensures that 
completions exist.
\end{example}

While the axioms for a causal site were modelled on the example where
regions are bounded subsets of a causally well-behaved Lorentzian
manifold, and 
$A \prec B$ iff every element of $A$ is in the causal past of every
element of $B$, we can use the structure we have to define a weaker
relation.

\begin{defn}
For regions $A$ and $B$, write $A \precsim B$ if there exist
non-empty subsets $A' \sse A$ and $B' \sse B$ with $A' \prec B'$
or $A' = B'$.
We say that $A$ \dfn{weakly precedes} $B$.
\end{defn}

In the example of a Lorentzian manifold (Example~\ref{ex:lorentz}), 
$A \precsim B$ iff some point of $A$ is in the causal past of some
point of $B$.

Note that $\precsim$ is not transitive.  That is, $A \precsim B$
and $B \precsim C$ does not imply $A \precsim C$.

\begin{defn}
For regions $A$ and $B$ with $A \prec B$, we say that $C$ is
\dfn{strongly complete} for $A$ and $B$ if for all non-empty 
$A' \sse A$ and $B' \sse B$, any causal path from $A'$ to $B'$ can
refined to a causal path from $A'$ to $B'$ one of whose members is
contained in $C$.
\end{defn}

Note that we do not require that $A \prec C \prec B$, since a strong
completion will rarely satisfy this.
A strong completion is ``wider'' than an ordinary completion.
See Figure~\ref{fig:completions}.

\begin{defn}
A region $B$ is a \dfn{cover} of a region $A$ if $B \prec A$
and if every causal path ending in $A$ can be refined to a
causal path ending in $A$ one of whose elements is contained in $B$.

A finite set of regions $\{B_i\}$ is a \dfn{cover} of $A$
if the union is a cover of $A$.
\end{defn}

\begin{remark}
We can regard a causal site as a category whose objects are the regions
by saying that there is one morphism from $A$ to $B$ if $A \prec B$ or
$A = B$, and no morphisms otherwise.
The collection of covers defined above has the feel of being a
Grothendieck topology~\cite{MM}.
But even if pullbacks exist in this category (they do, for example,
in the case of Example~\ref{ex:lorentz}), the pullback of a cover
is not necessarily a cover.
\end{remark}

Below we shall define a 2-category whose 1-arrows are all composable
strings of arrows in this category, i.e.\ causal paths.

\begin{defn}
A causal site is \dfn{Noetherian} if every chain of strict 
descending regions $A_1 \sps A_2 \sps A_3 \sps \cdots$
is finite. It is \dfn{locally Noetherian} if every 
local site is Noetherian.
\end{defn}

Noetherian implies locally Noetherian.

The causal site $\finite(M)$ described in Example~\ref{ex:finite}
is always Noetherian.
If $M = \N$ with the usual order, then the causal site
$\power(M)$ described in Example~\ref{ex:power} is locally
Noetherian but not Noetherian.

\begin{defn} 
A non-empty region $A$ in a causal site is an \dfn{absolute point}
if it contains no subregions besides $\phi$ and $A$.
If $A$ and $B$ are non-empty regions in a causal site, we say that $A$ is a 
\dfn{relative point} for $B$ if $A \prec B$ and for any $C$ with $C \prec B$
either $A = A_C$ (i.e.\ $A \prec C$) or $A_C = \phi$.
\end{defn}

If $A$ is an absolute point and $A \prec B$, then $A$ is a relative
point for $B$. 
And if $A$ is a relative point for $B$, then any non-empty subregion
of $A$ is also a relative point for $B$.

In Example~\ref{ex:lorentz}, an absolute point is just a point in
the usual sense, and every relative point is an absolute point.

In Examples~\ref{ex:power} and~\ref{ex:finite}, the absolute points
are just the elements of the poset, but the relative points can be
larger.  For example,
let $Q$ be the poset $\{a, b, b', c\}$ with $a < b < c$, $a < b' < c$
and with $b$ and $b'$ unrelated.  Then in the causal site $\power(Q)$,
the subset $B = \{b, b'\}$ is a relative point for $C = \{c\}$.
It is a union of the absolute points $\{b\}$ and $\{b'\}$.

We believe that because of the theorems that only a finite dimensional
space of information can flow across
horizons in general relativity~\cite{FMW} there should be 
causal sites associated to Lorentzian manifolds which have
interesting examples of relative points.  The information contained in the
relative position of a finite set of regions should exhaust what an
external observer can see. 

\begin{defn}
If $A$ and $B$ are regions in a causal site, we say that $A$ is 
\dfn{amply pointed}
with respect to $B$ if $A \prec B$ and if $C, D \sse A$ then either $C$ or 
$D$ contains a relative point for $B$ which the other does not, or $C = D$.
A causal site is \dfn{amply pointed} if all causal pairs in it are.
\end{defn}

\begin{defn}\label{de:grained}
A causal site which is locally Noetherian and amply pointed
is a \dfn{grained world}.
\end{defn}
In addition to the examples coming from causal sets, we believe that the family of
Noetherian causal sites described in Section~\ref{se:geometry} produces
grained worlds whenever there are no pairs of points connected by
more than one Planck-scale geodesic.
This still needs proof.

Although one can think of the regions in a causal site as like 
the bounded subsets of
a causal spacetime, we believe the interesting cases will be very
different from such classical examples. The Noetherian property
discussed above is an approach to imposing a Planck scale cutoff on
the structure of a causal site.

\section{The Intrinsic Geometry of Causal Sites}
\label{se:geometry}

It turns out that a causal site can contain more information than
just the causal structure of a manifold.  The reason 
is that a causal site may have a fundamental graininess which
sets a length and time scale. 
Physically, this graininess is expected to occur at 
the Planck scale, and serves as a measuring rod or clock.
Heuristically, a measurement at a smaller scale would result in
the formation of black hole, so the maximum possible number of
successive measurements along a timelike path gives its duration in
Planck units.

\begin{defn}
A causal site is \dfn{causally finite} if for any two regions $A \prec B$,
the length of any causal path from $A$ to $B$ is bounded above by a
constant $M_{A,B}$.
\end{defn}

We now want to interpret the least upper bound for the length of a causal
path between two causally related regions as the discretized timelike 
separation between them in Planck units. This is quite similar to the notion 
of length in a discrete causal set in~\cite{So}.

We now give examples to show that the resulting geometry can be quite 
interesting.

\begin{example}\label{ex:minkowski-diamond}
Consider Minkowski space.
Recall from Example~\ref{ex:lorentz} that a \dfn{diamond} is a region
$D(p,r) = \{q \st p \ll q \ll r\}$.
Define a \dfn{fundamental diamond} to be a region $D(p,r)$
where the proper time from $p$ to $r$ is 1.
Consider the set of bounded regions
which are unions of fundamental diamonds.
Define $\sse$ and $\prec$ as in Example~\ref{ex:lorentz}.
This forms a causal site.

Let's check the non-trivial axioms.
For the cutting axiom (Axiom~\ref{ax:cutting}),
let $B_A$ be the union of all fundamental diamonds $D$
such that $D \prec A$ and $D \sse B$.
This is bounded (contained in a finite union of diamonds) since $B$ is.
Note that it can happen that $B_A$ is empty even though there
are points of $B$ which are in the past of $A$.

For the completion axiom (Axiom~\ref{ax:completion} and
Definition~\ref{de:completion}),
a completion of $A \prec C$ can be taken to be the union of all
fundamental diamonds $D$ such that $A \prec D \prec C$.
This is bounded because it is a subset of the diamond
$D(a,c)$ for any points $a \in A$ and $c \in C$.

The remaining axioms are straightforward.
\end{example}

\begin{figure}
\centering
\epsfig{file=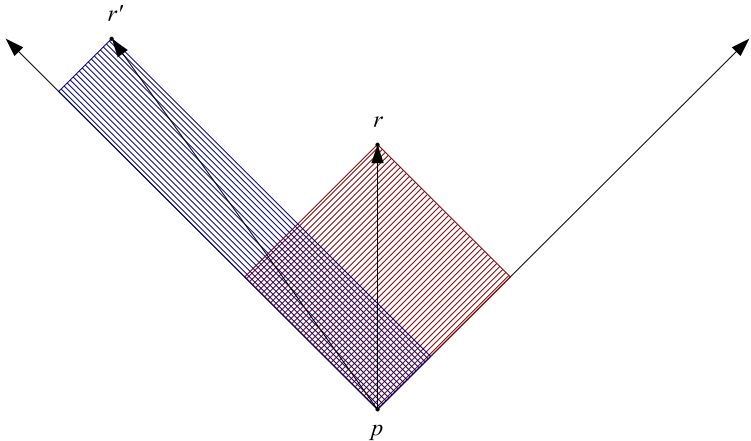,width=4in}
\caption{Fundamental diamonds}
\label{fig:diamonds}
\end{figure}

Figure~\ref{fig:diamonds} gives two examples of fundamental diamonds.  
The timelike vectors shown have proper length $1$.

The following theorem explains how this causal site
captures the geometry of Minkowski space.

\begin{thm}
Let $\gamma$ be a finite timelike geodesic segment in Minkowski space,
starting at the point $s$ and ending at $t$.
In the causal site of Example~\ref{ex:minkowski-diamond},
consider the causal paths $A_1 \prec \cdots \prec A_m$ 
with $s \in A_1$ and $t \in A_m$.
Then the proper length of $\gamma$, rounded up to an integer,
is equal to
the least upper bound of the lengths of these causal paths.
\end{thm}

The key element of the proof is the observation that using fundamental 
diamonds in any other rest frame than the one determined by the geodesic
produces a shorter chain. This is clear from the above picture.
In physical applications, we would use units in which the Planck time is $1$.

To handle curved spacetimes, we need to introduce the concept of
stable causality.
A Lorentzian manifold $M$ is \dfn{stably causal} if it has
no closed timelike curves, and every small perturbation of
the metric $g$ also has this property.  This can be expressed
by saying that there exists a metric $h$ whose light cones are
wider than those of $g$ and such that $h$ has no closed timelike
curves.  This condition is more physical that simply requiring
no closed timelike curves, since measurements have only finite
accuracy.

It is a non-trivial result \cite[Prop.~6.4.9]{HE}
that stable causality is equivalent to the existence of a global
time function.
In particular, a stably causal spacetime is automatically
time orientable.

Suppose that $M$ is stably causal and we have chosen a time orientation.
It is shown in~\cite{Pen} that since $M$ is stably causal,
the diamonds $D(p,r)$ determine the topology of $M$.

\begin{example}\label{ex:lorentz-diamond}
Generalizing the previous example, 
let $M$ be a stably causal Lorentzian manifold.
Define a \dfn{fundamental diamond} to be a region $D(p,r)$
such that there is a future-directed timelike geodesic of
length $1$ from $p$ to $r$.
Consider the set of bounded regions
which are unions of fundamental diamonds.
Define $\sse$ and $\prec$ as in Example~\ref{ex:lorentz}.
This forms a causal site.
The proof is the same as Example~\ref{ex:minkowski-diamond}.
\end{example}

The following conjecture explains how these causal sites
capture the large-scale geometric information of the manifold.

\begin{conjecture}
Let $M$ be a stably causal Lorentzian manifold
whose sectional curvatures are much less than $1$,
and let $\gamma$ be a timelike curve from $s$ to $t$ in $M$ 
whose radius of curvature and length are large.
In the causal site of Example~\ref{ex:lorentz-diamond},
consider the causal paths $A_1 \prec \cdots \prec A_m$ 
such that each region $A_i$ intersects $\gamma$.
Then the proper length of $\gamma$ is well approximated by
the least upper bound of the lengths of these causal paths, with an error 
which is small compared to the length of $\gamma$.
\end{conjecture}

Of course, the precise statement of the error of the approximation in terms
of the curvatures will require a more delicate analysis.

We believe this causal site reflects classical geometry 
more accurately than a causal set can.
This is because we have enough relative points 
to adjust to any direction to get the best fit. 

It is well known that the lengths of timelike curves (clock times), are 
enough to describe the geometry of a Lorentzian manifold completely.
Thus our conjecture implies that any solution to general relativity
whose sectional curvature is small compared to the Planck scale, 
with any type of matter whatsoever, 
can be approximated by the intrinsic geometry of a causally finite 
causal site with accuracy on the order of the Planck scale.

The causal sites discussed in this section have a fundamental
graininess, but are nevertheless not Noetherian or locally Noetherian.
It is not hard to invent modifications of our examples which would have 
stronger finiteness conditions and still recover interesting 
geometrodynamics.
For example, fix a discrete closed subspace $L$ of Minkowski space and
a discrete closed subspace $S$ of the hyperboloid of unit timelike
vectors.  Now consider bounded regions $R$ in Minkowski space 
which are unions of
fundamental diamonds $D(l-s/2,l+s/2)$ centered on points $l$ of $L$
and pointing in directions $s$ in $S$.
Since $R$ is bounded, it contains
only finitely many points in $L$.  Moreover, in any sufficiently
boosted coordinate system, it would be too contracted to contain a
fundamental diamond.  Thus it only contains finitely many fundamental
diamonds whose directions are contained in $S$ and which are centered
on points of $L$.  Therefore this collection of regions forms a
Noetherian causal site which reproduces the spacetime geometry of
Minkowski space at a large scale. Such a model is computationally
accessible, and is a good candidate for quantization, as discussed below.

The discrete subspace $L$ of Minkowski space could be chosen to be
a random sprinkling of points, or a lattice.
In a general manifold, $L$ would have to be an irregular collection
of points.  If the manifold is globally hyperbolic, then one again
obtains a Noetherian causal site.

Note that the use of fundamental diamonds is not crucial to any of the
discussion in this section.  One could use other shapes, such as
``fundamental cylinders,'' which are regions formed by working in
normal coordinates and taking the cartesian product of a timelike
interval of length 1 with a ball of radius 1 in the hyperplane
orthogonal to the chosen timelike direction.

As yet, we do not know how to impose Einstein's equation on a causal site 
purely intrinsically.

\section{Quantum Sites}

A causal site can be thought of as a large number of answers to questions, 
either telling
us one region is inside another, or that one region can observe another.

The answers to these questions can be grouped together in certain good
examples and reinterpreted as describing physically interesting
geometries.

It is therefore natural to reinterpret the statements that define a 
causal site as quantum observables and attempt to extract a quantum geometry 
from them. This could be a new avenue of attack on the problem of quantizing
gravity.

It is not hard to see how to begin such a program. We could tensor together a
finite dimensional Hilbert space for each pair of regions, and construct 
operators for containment and causal relatedness on each. The interesting 
part would be to see if suitable commutation relations could be found on the 
operators to reproduce Einstein's equation in the classical limit, for 
suitable families of causal sites. We have not yet investigated this.

\section{Simplicial Structures and Causal Sites}
\label{se:simplicial}

We first remind the reader of the relationship between partial orders
and simplicial sets.  A good reference for simplicial sets is~\cite{GJ}.

Roughly speaking, 
a simplicial set is a set of abstract simplices (points, edges, triangles,
tetrahedra, etc.)\ with the faces of the simplices of dimension $n$
identified with simplices of dimension $n-1$, thus gluing the simplices
together into a combinatorial model of a space. The mathematically
natural definition of a simplex includes an ordering of its vertices,
and includes ``degenerate'' simplices in which some vertices are
repeated.
If $X$ is a simplicial set, we write $X_n$ for the set of $n$-simplices.

Associated to a given partially ordered set is a simplicial set
which contains all of the information.
An $n$-simplex of this simplicial set is a weakly ascending chain of 
length $n+1$.
The face maps come from omitting one member of the chain,
and the degeneracies from repeating one member.

Since the regions of a causal site have two partial orders on them,
they have a natural description as the vertices of a bisimplicial
set. Roughly speaking,
a bisimplicial set is a collection of abstract cartesian products of pairs
of simplices, with attachments along face maps corresponding to both
simplicial factors separately. The product of a triangle with a
tetrahedron, for example, has three faces which are
edge $\times$ tetrahedron, and four which were triangle $\times$ triangle.
If $X$ is a bisimplicial set, we write $X_{m,n}$ for the set of
$(m,n)$-bisimplices.

Let us give an explicit description of the bisimplicial set of a
causal site.
An $(m,n)$-bisimplex is a family $A_{i,j}$ of regions in the
site, $i=0,\ldots,m$, $j=0,\ldots,n$, such that 
if $a<b$ then $A_{a,j} \prec A_{b, j}$ or $A_{a,j} = A_{b, j}$
and if $c<d$ then $A_{i,c} \sse A_{i,d}$. 
The two types of face maps come from omitting one value of $i$ or $j$,
and the two types of degeneracies from repeating items in the sequences.

\begin{defn}\label{de:elem-class-space}
The bisimplicial set associated to a causal site above is called its 
\dfn{elementary classifying space}.
\end{defn}

Note that if $X$ is any bisimplicial set, and we fix $m$, then
there is a natural simplicial set $X_{m,\cdot}$ whose $n$-simplices are the
$(m,n)$-bisimplices of $X$.  We call $X_{m,\cdot}$ the 
\mdfn{simplicial set of $m$-simplices}.

It is now an important observation that the elementary classifying space
of a causal site satisfies the 
\dfn{Segal condition}~\cite[Defn. Ta]{Le}, \cite{Ta}. The Segal condition for a
bisimplicial set has two parts. Translated into our situation, the
first part states that the simplicial set of $m$-simplices
is the subset of the product of $m$ copies of
the simplicial set of $1$-simplices, where the
adjacent $0$-simplex objects are equal.
More explicitly, it says that for each $m$ and $n$, the natural map
\[
  X_{m,n} \ra X_{m,1} \times_{X_{m,0}} \cdots \times_{X_{m,0}} X_{m,1}
\]
must be a bijection. 
This is true for the elementary classifying space, essentially
because giving a chain $A_{i,0} \sse \cdots \sse A_{i,n}$
is the same as giving $n$ chains $A_{i,0} \sse A_{i,1}$,
$A_{i,1} \sse A_{i,2}$, \dots, $A_{i,n-1} \sse A_{i,n}$.

The second part of the Segal condition for bisimplicial
sets is more subtle in general.
The first part of the Segal condition tells us that for
each $m$, the simplicial set $X_{m,\cdot}$ can be regarded
as a category.  The second part of the Segal condition
then requires that for each $m$, the natural map
\[
  X_{m,\cdot} \ra X_{1,\cdot} \times_{X_{0,\cdot}} \cdots \times_{X_{0,\cdot}} X_{1,\cdot}
\]
must be an equivalence of categories.
In our case, something stronger is true:  for each $m$ and $n$, the natural map
\[
  X_{m,n} \ra X_{1,n} \times_{X_{0,n}} \cdots \times_{X_{0,n}} X_{1,n}
\]
is a bijection.
The reason this is true is similar to the reason that the first
part of the Segal condition holds.
And from this it follows that the categories above are equivalent.

Now let us describe 
the \dfn{Segal condition} for a simplicial set $A$.  It simply states that
there is exactly one $n$-simplex for each chain of $n$ $1$-simplices
with the second vertex of the $i$th $1$-simplex equal to the first
vertex of the $i+1$st.  That is, it states that the natural map
\[
  A_n \ra A_1 \times_{A_0} A_1 \cdots \times_{A_0} A_1
\]
is a bijection.
This is true, for example, for the simplicial set associated to
any partially ordered set.
We will explain the importance of this condition below, when we study
the relationship with category theory.

One point of all this is that we can completely capture the structure
of a causal site as a combinatorially described space. This allows us
to translate the problem of constructing physical theories on a causal
site into the construction of presheaves over such a space, a problem which
is well understood. Another is to join the concept of causal sites to
the field of higher category theory.

Now we need to construct a modification of the elementary classifying
space of a causal site. The modification will allow us to associate to
a causal site a ``special'' bisimplicial set; 
see~\cite{Ta}, where it is referred to as axiom C1, and~\cite[Defn. Ta]{Le}.

\begin{defn}
A bisimplicial set $X$ is \dfn{special} if the simplicial
set $X_{0,\cdot}$ of $0$-simplices is a disjoint union of points.
In other words, all of the face and degeneracy maps between sets
of the form $X_{0,n}$ are bijections.
\end{defn}

This condition is important in category theory, because it is
necessary for a bisimplicial set to be the nerve of a weak
2-category. As we shall see, this condition also plays a natural role
when we try to construct models for quantum physical systems.

Now we want to define a new version of the classifying space of a
causal site.

\begin{defn}
The \dfn{physical classifying space} of a causal site is the
bisimplicial set whose $(m,n)$-bisimplices correspond to the
following data:
\begin{enuma}
\item a sequence $A_0, \ldots, A_m$ of regions with $A_i \prec A_{i+1}$;
\item for each $i=0, \ldots, m-1$ and $j=0, \ldots, n$ a causal path 
      $C^{j}_i$ from $A_i$ to $A_{i+1}$;
\end{enuma}
such that for each $i$ and $j$, the causal path $C^j_i$ is a refinement of
$C^{j+1}_i$.
The two types of face maps are given by composition of rows and columns 
respectively.
\end{defn}

In this definition, refinements of causal paths play a central role. 
The physical motivation
for this is the goal of writing discrete analogs of Feynman path
integrals, by summing effects of propagation along causal paths with
the same beginning and end. Mathematically, we will see below that the
$(m,n)$-bisimplices of the physical classifying space fit together to
form a weak 2-category, because the initial and final paths of a
refinement share a common initial and final region.

We would like to formulate the idea that all the information of a
causal site is contained in its physical classifying space. We
therefore make the following tentative definition:

\begin{defn}
A causal site is \dfn{tractable} if its elementary and physical 
classifying spaces are homotopy equivalent.
\end{defn}

We do not know which causal sites are tractable, or whether this will
prove to be a useful way to formulate the idea that the physical
classifying space ``suffices''.

\begin{prop}
The physical classifying space of a causal site is a special 
bisimplicial set satisfying the Segal condition.
\end{prop}

The proof is similar to the argument given for the elementary
classifying space, including the fact that a stronger form
of the second part of the Segal condition holds.

Thus, as we explain below,  the physically natural idea of focussing
on the causal paths of a
causal site supplies the missing mathematical ingredient to replace a
spacetime with a weak 2-category. We think this is suggestive for our
program of applying higher categorical ideas of topology to quantum physics.

\section{The Weak 2-category Structure of a Causal Site}
\label{se:2-cat}

There is a very strong connection between simplicial and
multisimplicial sets and higher categories.  We begin
with a standard theorem about (1-)categories \cite[p.\ 34]{Le}, 
\cite{Ta}.

\begin{thm}
There is a one-to-one correspondence between categories and simplicial
sets satisfying the Segal condition.
\end{thm}

The connection can be described by taking the nerve of the
category. This is a simplicial set with one $n$-simplex for each string
of $n$ composable morphisms in the category, with face maps given by
compositions of pairs of adjacent morphisms and degeneracies given
by inserting identity maps.

This theorem has recently been extended from categories to
2-categories by Tamsamani \cite[Defn. Ta]{Le}, \cite{Ta}:  he
proved that there is a natural one-to-one correspondence between
special bisimplicial sets satisfying the Segal condition and weak
2-categories.

The connection proceeds by taking the ``2-nerve'' of the 2-category. 
This is the bisimplicial set with $(0,0)$-bisimplices
the objects of the 2-category, 
$(1,0)$-bisimplices the 1-morphisms in the 2-category, 
$(1,1)$-bisimplices the 2-morphisms in the 2-category.
A general $(m,n)$-bisimplex is a doubly indexed array of $mn$ 2-morphisms
along with some additional data.

The analogy with the structure of
the physical classifying space prompts the following definition:

\begin{defn}
If the causal path 
$A \prec P_1 \prec P_2 \prec \cdots \prec P_p \prec B$ 
is a refinement of the causal path
$A \prec Q_1 \prec Q_2 \prec \cdots \prec Q_q \prec B$
then we say that there is a \dfn{chain inclusion} from
the causal path $\{Q_j\}$ to the causal path $\{P_i\}$.

The \dfn{inclusion 2-category} of a causal site is the 2-category whose
objects are regions, 1-morphisms are causal paths, and 2-morphisms are
chain inclusions.
\end{defn}

The inclusion 2-category of a causal site connects regions, processes
between regions, and inclusions of processes. These are the elements
which would go into a description of an experiment. A 2-functor on this
2-category would then give us a mathematical language in which to
associate calculations to an experiment, by associating concrete
mathematical structures and relations to the objects and processes.

As we shall discuss below, the state sum models for quantum gravity
have a natural formulation in terms of 1- or 2-categories.  Any two
2-categories are connected by a 2-category of functors, natural
transformations and modifications. This means the categorical
structure we have described gives us a sort of calculus for
constructing physical models over causal sites using the state sum
models as local physical data.

To put things simply, one could easily feel that passing from point
sets to regions led to a mathematical wilderness. The structure we
have discovered has geometric, categorical and algebraic aspects
which make available to us a large framework of definitions and
theorems which can guide us to natural constructions of physical
models. The natural definition of a physical model on a causal site is
a type of functor on it. The definition could in principle be found
without the categorical language, but it is a foolish cave explorer
who throws away a light.

\section{Bisimplicial patches }\label{se:patches}

The geometrical and physical applications of smooth manifolds largely
grow out of the fact that to every smooth manifold we can naturally
associate a tangent bundle. This develops from the more elementary
fact that every point of a manifold has a neighborhood which can be
described as a space of a fixed and familiar kind. In this section we
propose such a neighborhood structure.
In the next, we show how to construct an
analog of the tangent bundle, but a relational, or 2-categorical one.

\begin{defn}
If $A$ is a region in a causal site, the bisimplicial set of 
regions contained in it is its \dfn{bisimplicial patch}.
This bisimplicial set sits naturally inside the elementary
classifying space (Definition~\ref{de:elem-class-space}).
\end{defn}

With this definition in place,
it is possible to treat causal sites in a manner analogous to
manifolds, by working on the local bisimplicial patches thought of as
coordinate patches.

Simplicial and bisimplicial sets have a homotopy theory equivalent to
the homotopy theory of topological spaces.  This subtle mathematical fact
should allow us to use the combinatorics of the bisimplicial patch as
an approximation to the local topology of a region.


\begin{defn}
Let $R \sse R'$ be regions.  The \dfn{bisimplicial patch}
of the complement is the bisimplicial set of regions $A$
which are contained in $R'$ and disjoint from $R$.
The \dfn{relative homotopy type} of $R$ in $R'$ means
the homotopy type of the pair $(X,Y)$ where $X$ is
the bisimplicial patch of $R'$ and $Y$ is the bisimplicial
patch of the complement of $R$ in $R'$.
\end{defn}

\begin{defn}
A region $R$ is an \mdfn{$n$-ball} if it is contained in a
region $R'$ such that the relative homotopy type is that
of an $n$-sphere.

A causal site is a \mdfn{grained $n$-manifold} if every
region is contained in a union of a finite number of $n$-balls.
\end{defn}

\section{The Tangent 2-Bundle}\label{se:2-bundle}

In a physical model on a causal site, we would like to think of
information being transferred along causal paths. We will now describe
a construction of a version of a tangent bundle for causal sites in
which the regions contain just such information as can be observed
along the causal paths, while the compositions and inclusions of
causal paths have a natural action.

\subsection{Causal paths}

For $A \prec B$ we define $\Path(A,B)$ to be the set of causal paths 
from $A$ to $B$.
For any $A \prec B \prec C$ there is a natural composition map 
$\Path(A,B) \times \Path(B,C) \ra \Path(A,C)$
which sends $A \prec P_1 \prec \cdots P_m \prec B$ 
and         $B \prec Q_1 \prec \cdots Q_n \prec C$ 
to $A \prec P_1 \prec \cdots P_m \prec B \prec Q_1 \prec \cdots Q_n \prec C$.
The composition map is injective.

For a region $A$, define the \mdfn{future cone of $A$} to be 
$\future{A} = \{ B \st A \prec B \}$ and define the
\mdfn{future tangent space of $A$} to be
$\subfuture{A} = \{ \future{A'} \st A' \sse A \}$,
the poset of future cones of subregions of $A$, ordered
by inclusion.
The future cone $\future{A}$ is a minimum element of $\subfuture{A}$.
For a region $B$, define the \mdfn{past cone of $B$} to be 
$\past{B} = \{ A \st A \prec B \}$ and define the
\mdfn{past tangent space of $B$} to be
$\subpast{B} = \{ \past{B'} \st B' \sse B \}$, ordered by inclusion.
The past cone $\past{B}$ is a minimum element of $\subpast{B}$.

The relevance of these cones is that if $\future{A_1} = \future{A_2}$, 
then $\Path(A_1,B) = \Path(A_2,B)$ for any $B$.
(This is an equality rather than a bijection because a path from
$A$ to $B$ doesn't include the regions $A$ and $B$.)
More generally, if $\future{A_1} \sse \future{A_2}$, 
then $\Path(A_1,B) \sse \Path(A_2,B)$ for any $B$.
Similarly, $\past{B_1} \sse \past{B_2}$ implies that
$\Path(A,B_1) \sse \Path(A,B_2)$ for any $A$.
Writing $\Fut$ for the poset $\{ \future{A} \}$ of all future cones
and $\Past$ for the poset $\{ \past{B} \}$ of all past cones,
this says that $\Path(-,-)$ is an order preserving map from the
poset $\Fut \times \Past$
to the poset $\Inc$ of sets and inclusions.
In fancier language, $\Path(-,-)$ is a functor.

This has various consequences.  For example, if
$A' \sse A$, then $\future{A'} \spse \future{A}$
and so $\Path(A',B) \spse \Path(A,B)$ for any $B$.
And if $A' \prec A$, then $\future{A'} \spse \future{A}$
and so again $\Path(A',B) \spse \Path(A,B)$ for any $B$.
Similar reasoning shows that if 
$B' \sse B$ or $B' \succ B$ (note the reversal!), 
then $\future{B'} \spse \future{B}$
and so $\Path(A,B') \spse \Path(A,B)$ for any $A$.

\subsection{Causal paths subordinate to a given path}

If we are given a causal path $P$ from $A$ to $B$, 
we can assign to any pair of subregions $A' \sse A$ and $B' \sse B$ 
the set $\Path_P(A', B')$ of causal paths from $A'$ to $B'$ which 
are refinements of $P$.
As above, this defines a functor, in this case from
$\subfuture{A} \times \subpast{B}$ to $\Inc$.

We next describe a 2-category which is a natural target for this
construction.

\newcommand{\pp}{\mathcal{PP}}
\begin{defn}
Define a weak 2-category $\pp$ in the following way.
The objects are pairs $(F,P)$ of posets with minimum elements
$m_F$ and $m_P$.
A 1-morphism from $(F_1,P_1)$ to $(F_2,P_2)$ is a functor
from $F_1 \times P_2$ to $\Inc$.
The composite of $S : F_1 \times P_2 \ra \Inc$ with
$T : F_2 \times P_3 \ra \Inc$ is the functor
$TS : F_1 \times P_3 \ra \Inc$ defined by
$TS(f,p) = S(f,m_{P_2}) \times T(m_{F_2},p)$.
If $S$ and $T$ are 1-morphisms from $(F_1,P_1)$ to $(F_2,P_2)$,
i.e.\ functors from $F_1 \times P_2$ to $\Inc$,
a 2-morphism from $S$ to $T$ is a natural family of
injections $S(f,p) \ra T(f,p)$.
\end{defn}

Note that the composites $(TS)R$ and $T(SR)$ of 1-morphisms
are in general not equal.  Instead, there is a natural
bijection between them.  Thus $\pp$ is a \emph{weak} 2-category.
This also explains why the 2-morphisms allow arbitrary injections
rather than just inclusions.

The horizontal and vertical
compositions of the 2-morphisms of $\pp$ are given by cartesian product and
composition of injections, respectively. The coherence of $\pp$ is natural.

The discussion above can be summarized by saying that there
is a weak 2-functor from the inclusion 2-category of our
causal site to the weak 2-category $\pp$.
This can be thought of as a prestack.
It sends a region $A$ to the pair $(\subfuture{A}, \subpast{A})$
of posets.
It sends a causal path $P$ from $A$ to $B$ to the
functor $\Path_P(-,-)$ from $\subfuture{A} \times \subpast{B}$ to $\Inc$.
And it sends a chain inclusion from $P'$ to $P$
(which means that $P$ and $P'$ are both causal paths from
$A$ to $B$, and $P'$ is a refinement of $P$)
to the natural family of inclusions $\Path_{P'}(A',B') \sse \Path_P(A',B')$.
If $P$ is a causal path from $A$ to $B$
and $Q$ is a causal path from $B$ to $C$,
there is a natural injection 
$\Path_P(A',B) \times \Path_Q(B,C') \ra \Path_{Q\circ P}(A',C')$,
where $Q \circ P$ denotes the composite of $P$ and $Q$.

\begin{defn}
We call this weak 2-functor the \dfn{tangent 2-bundle} of the
causal site.
\end{defn}

The tangent 2-bundle contains information about how causal paths
can link the observable tangent spaces together.
Since we can also refine
causal paths, expansions into discretized path sums will be possible by
decomposing intermediate regions into unions and summing over the
families of refined paths which occur.

We think the analogy between smooth manifolds and causal sites may be
a good guide to applications. We use the tangent bundle as a setting
in order to apply calculus to the geometry of manifolds. Similarly,
the local simplicial or bisimplicial structures on a site may enable
us to apply the calculus of categorical state sums to quantum geometry
on them. In constructing a quantum theory over a causal site, it should be 
possible to use the connecting complexes as discretized analogs of a Feynman 
path integral, and their 2-categorical structure should help constrain such 
a construction, allowing us to find interesting models just from the 
requirement of 2-functoriality.

\section{Unitary Prestacks on a Causal Site}

In the next few sections, we investigate a more conventional approach to 
constructing geometrical or physical models on causal sites, namely, putting 
suitable presheaves and prestacks over them.

We will begin with a preliminary investigation of the type of physical
model which the mathematical structure of a causal site suggests. We
will not attempt to construct specific physically realistic models in this
paper.

The original idea of a site was motivated by the fact that a
presheaf over a topological space $X$ is equivalent to a contravariant
functor from the category of open subsets of $X$.

Since a bundle, including the tangent bundle of a manifold, can be
regarded as a type of sheaf, we have a language for describing analogs
of the basic structures underlying Yang-Mills theory and general
relativity available for sites, although we will probably need
stronger regularity assumptions to be able to create more precise
analogs.

It is interesting that including a causal structure in our model, which
traditionally is expressed by changing from a Riemannian to a Lorentzian
metric, is expressed here by passage from a 1-category to a
2-category, a process which has been called 
categorification~\cite{CF,BD}.

This has the immediate consequence that rather than looking for models
over a causal site within an ordinary category, such as the category
of Hilbert spaces, we must turn to a 2-category, such as the category
of 2-Hilbert spaces \cite{Ba1}. Instead of a presheaf of Hilbert spaces, 
we work with a prestack of 2-Hilbert spaces.
A prestack is a weak 2-functor from our causal site to a 
2-category~\cite{Br}.
In a prestack, a triangle of restriction morphisms only needs to commute
up to a 2-morphism.

This means that rather than thinking of an assignment of a single
Hilbert space to a region (something like the space of quantum states
propagating through it), we assign to it a category of Hilbert
spaces. Physically, we can think of this as related to the idea that
the state space of a region has dimension related to the area of its
boundary, so that if the geometry is itself a quantum variable, then
the Hilbert space itself cannot be unique.

Indeed, if we tried to construct a physical theory over a causal site,
we would quickly discover that we couldn't find a natural linear map
to associate to containment, because of the nonlocal correlations in
quantum theories. We would be trapped trying to map pure states to
mixed states.

It actually provides an interesting new slant on the interpretation of
quantum mechanics rewriting this process in terms of the different
choices of Hilbert spaces in the categories corresponding to two
regions, one of which contains the other.

Let us explain in a very simple situation, using the category
2-Vect \cite{KV} rather than 2-Hilb, how this would look. Objects in
2-Vect are $n$-vector spaces, i.e.\ the category of all $n$-tuples of
vector spaces for some $n$. Arrows in 2-Vect are functors
defined by tensoring by some column of vector spaces. Thus, a
(contravariant) functor from a causal site to 2-Vect would find the
vector spaces which could appear on a subregion being combined into
tensor products of vector spaces associated to a larger region.

Now there is neither a natural map $A \rightarrow A\otimes B$ nor 
$A \otimes B \rightarrow A$, so there would be no way to represent
inclusion as a map on individual Hilbert spaces.  This is a
mathematically elegant way to express the existence of nonlocal
correlations in compound systems in quantum mechanics. The overall
rule which tells us how the different Hilbert spaces associated to
regions are related has an elegant functorial form, which cannot be
recovered on the individual Hilbert spaces. In \cite{HMS}, a similar
problem was studied for models on causal sets, and it was discovered
that maps between causally related points had to be expressed as
completely positive maps on spaces of mixed states (i.e., Hermitian
operators over the Hilbert spaces, rather than the Hilbert spaces
themselves). The setting we are exploring, combining both causality
and containment, should contain examples of this, but set in a broader
algebraic context.

Now we can ask ourselves which of the models which have appeared in
mathematical physics would be good candidates for extending to a
causal site. An obvious choice would be 2-Yang Mills 
theory~\cite{Ba2,GP}, which is a theory which puts a nonlinear partial
differential operator on a 2-form, by analogy with the Yang-Mills
equation on a 1-form.

Since 2-Yang Mills theory has the interpretation of physics on the
curvature of a gerbe, which is really a special type of 
stack of categories, it will
have a direct generalization to causal sites.

\section{A Brief Review of Categorical State Sum Models}

Categorical state sums are models of
quantum geometry on simplicial complexes\linebreak\cite{CKY, BC, CY, NR, CPR}. 
They are constructed out of
unitary representations of Lie groups or quantum groups, intertwining
maps of these representations, and in some cases, 2-intertwiners. The
maps are composed in closed patterns related to the combinatorics of
the simplicial complex, traced, and summed over.  The result is a
discrete version of a path integral for quantum geometry, with the
unitary representations acting as Hilbert spaces on which geometric
quantities are represented as operators.

For example, in the model  of \cite{BC}, we have a four-dimensional
simplicial complex. The 2-simplices (triangles) are labelled with
unitary representations of the Lorentz group.  We think of these as
quantizations of the oriented area elements a geometry would assign to
the faces, because oriented area elements can be identified with
elements of the Lie algebra of the Lorentz group.

In fact, in order to impose the necessary constraints on the geometry,
we restrict ourselves to a class of representations called the
balanced ones.  We then label the tetrahedra with a sum over the
balanced intertwining maps between the four representations, trace
over each 4-simplex, multiply, then sum over all possible choices of
representations.

There are a number of models of this type, differing by the dimension,
choice of symmetry group, constraints, etc. These amount to different
choices of the geometry to be quantized.

This type of model is closely related to the ideas discussed above
describing simplicial sets as categories. They can be written as
sums of functors from the category associated to the complex to the
category of representations of the geometrical symmetry group or
quantum group.

In short, the categorical state sum picture makes it natural to assign
a quantum geometry to a simplicial complex.

In the context of causal sites, this can be thought of as a candidate
for local quantum geometry on a tangent space, analogously to the use
of inner products on tangent spaces in classical Riemannian geometry.

One of the reasons that manifold theory was so important to the
development of relativity was that it gives a natural concrete
expression for the equivalence principle: laws must be expressed in
tensor form. The categorical underpinning of the structure of causal
sites suggests a similar principle in quantum gravity: laws must be
expressed in functorial form. This simple and attractive principle
will cut down the possibilities greatly.

Given that we already have subcategories which express the relational
world seen by an observer, it is also a natural expression of the idea
behind the equivalence principle, if we take the point of view that
coordinate systems are idealized observers.

\section{State Sum Models on Causal Sites}

The physically interesting examples of Section~\ref{se:geometry} 
all possess local symmetries.
(Since a bounded region which is boosted too far no longer contains 
fundamental cylinders, a mathematical formalization of this would need to 
use actions of germs of the Lorentz or Poincare groups. We will not work 
this out in this paper.)

It therefore seems that the Hilbert spaces in a quantum site would group 
together in unitary representations of the Lorentz group. 

Now one of the great challenges for our proposal is to find a way to derive
Einstein's equation as a classical limit for some formulation of the 
dynamics on a quantum site. Categorical state sum models are interesting, because the
laws of combination of the representations produce a classical limit which 
implies Einstein's equation. In other words, Einstein's equation is actually 
determined by the symmetry of the model, expressed in the tensor category of 
unitary representations of the Lorentz group.

Now is it possible that we can introduce categorical state sum models 
into a suitable family
of causal sites with local symmetries? Let us try to imagine how that could 
arise.

In order to define a class of causal sites which would be physically
realistic near the Planck scale, we have defined ``grained worlds''
above as locally Noetherian, amply pointed causal sites. We think of
this as a way to embed facts about the limited amount of information
which can flow between regions in general relativity into the fabric
of spacetime itself.

The combination of these two properties means that given a causally
related pair of regions $A \prec B$, we can find a finite set of
relational points for $B$ in $A$ which in some sense exhaust the
information which $B$ can see in $A$. This would mean that the
relational tangent space would be adequately described by a simplicial
set, assuming the set of points did not contain one another.

This suggests that the state sum model of \cite{BC}, or some similar model, might
appear as a good approximation to the quantum theory of the part of a 
quantum site contained in a region. 

This is probably a better setting for state sum models than direct application to 
a whole universe, since the finiteness of the simplicial set becomes better 
motivated.

Then, when we wanted to express the flow of information from one region, 
described 
by a state sum on a simplicial set, to another, along a given
causal path, we would know that the map which described it
would have to behave well when we included one causal path in another.

This includes the information flow law in a 2-categorical structure, 
which might be rich enough to allow us to determine it.

So a program suggests itself, to make a suitable set of assumptions on
a prestack over a grained world, so that if we pass to a set of
relational points the relative position information ends up containing
all the transferred information, and takes the form of a state sum
model.

At this point, we realize that our new point of view is making us ask
a series of new questions about state sums which had no motivation in
the past when they were thought of as stand alone models. We need
to express the relationship between two state sum models when the
complex of one is embedded in the other, we need to work out
expressions for causal flows of information between state sums, and we
need to investigate in what sense they approximate points with quantum
variables between them as they grow far apart. If nothing else, the
causal sites proposal has widened our perspective on state sum models.

\section{Conclusions and Future Directions}

We now have defined a new class of mathematical objects, which are not
point sets, but which can function as a topological foundation for
classical general relativity. We have families of examples
corresponding to classical spacetimes, which include descriptions of
the geometrodynamics of classical spacetimes above the Planck scale,
without the inclusion of a metric tensor as a mathematical datum
separate from the underlying spacetime structure.

The representation of geometry by a causal site is analogous to the 
representation by a discrete causal set, but as we have discussed, it seems 
to give a much closer approximation.
Our relative and absolute points should not be confused with ordinary 
points, either in manifolds or causal sets. They are not ``atomic.''
For example,
a union of a finite family of fundamental diamonds in a Lorentzian manifold 
can contain 
an infinite number of other fundamental diamonds. Nevertheless, the causality 
relationship between them gives rise to a finiteness or discreteness in the 
structure. It therefore makes a bridge between continuous and discrete 
structures in a new way.
The hypothesis of overlapping minimal regions seems more physically natural 
to us than the discrete point sets of causal sets or categorical state sum 
models. 
It also means that the continuous or discrete symmetry 
properties of a causal site can be more robust than a causal set.
In the examples we have constructed, the symmetry with respect to the Lorentz
group is not broken, because we have fundamental diamonds in all reference 
frames. In the theory of causal sets, by contrast, it is only possible to 
recover the symmetry on the average in the infinite volume of Minkowski 
space~\cite{So}. The fact that we are able to produce a mathematical 
construction which combines a fundamental graininess with symmetry in local 
regions gives our construction added interest.

We think it will be interesting to try to find further families
of examples, for instance to try to construct quantum $n$-manifolds
neither related to causal sets nor directly derived from manifolds. Different
classes of these might give new examples of ``spacetime textures,''
which could be explored as new settings for physical models.
It may well be that the physical intuition expressed in the phrase
``spacetime foam'' can find better expression in our setting that
when constrained by the necessity of considering only families of
smooth manifolds.

Causality and time are integrated into the mathematical structure
of a causal site in a striking fashion: 
spacetime appears as a categorification of space.
The fact that causal sites can be considered either as bisimplicial
sets or as 2-categories means we have a number of well-understood
tools at our disposal for constructing examples.

In addition, the structure of a causal site seems to lend itself
readily to quantization. We have not yet examined this in any detail.

We have, at this point, many more questions than answers. Can the geometric 
picture of Example~\ref{ex:lorentz-diamond} be generalized to more intrinsic 
models, i.e.\ models which have definitions not so directly dependent
on a classical spacetime?
Can it be quantized? Since it seems natural to approach quantization
by beginning with the family of containments of a particular causal
site, it will be easier to find semiclassical states in this approach
than in other approaches to quantum gravity.

It would seem that the Hilbert spaces arising from quantization of
a causal site with local symmetries would decompose into unitary 
representations of the Lorentz group. Do the state sum models arise in 
such a picture? We should remind ourselves that Einstein's equation is almost 
determined by symmetry considerations. Would a quantization of a causal site 
with Lorentz symmetry which was required to respect the
symmetry be nearly determined as well? 

We also have not yet studied the description of curvature 
in causal sites, nor searched for a geometric way to impose Einstein's 
equation.  

General relativity is intimately bound up with our ideas about spacetime 
and geometry. Changing the foundation of these creates a situation in which 
all the questions of classical and quantum relativity can be reexamined. The 
large family of accessible examples we have constructed means that a fairly 
broad program can be practically implemented.

\newpage
\noindent
\textbf{Acknowledgments:} The authors wish to thank Rick Jardine and
Louis Kauffman for many helpful comments. The second author also
wishes to thank the Universit\'e de Montpellier 2 and the University
of Western Ontario for their hospitality, and Louis Kauffman for his
hospitality at the University of Waterloo.

The first author was supported by an NSERC Discovery Grant.

\end{document}